\documentclass{article}
\usepackage{ijcai26}

\usepackage{times}
\usepackage{multirow}
\usepackage{soul}
\usepackage{url}
\usepackage[hidelinks]{hyperref}
\usepackage[utf8]{inputenc}
\usepackage[small]{caption}
\usepackage{graphicx}
\usepackage{amsmath}
\usepackage{amsthm}
\usepackage{booktabs}
\usepackage[switch]{lineno}
\usepackage{algorithm}
\usepackage{algpseudocode}
\usepackage{amsmath, amssymb}
\usepackage{mathrsfs}
\usepackage{booktabs}
\usepackage{caption}   
\usepackage{subcaption}

\title{DTAMS: High-Capacity Generative Steganography via Dynamic Multi-Timestep Selection and Adaptive Deviation Mapping in Latent Diffusion}

\author{
Yuhao Xue$^1$\thanks{Equal contribution}
\and
Jiuan Zhou$^1$\footnotemark[1]
\and
Yu Cheng$^{1,2}$\footnotemark[1]
\and
Zhaoxia Yin$^{1}$\thanks{Corresponding author: zxzyin@cee.ecnu.edu.cn}
\\
\affiliations
$^1$East China Normal University\\
$^2$Shanghai Innovation Institute\\
\emails
10242140454@stu.ecnu.edu.cn\\
yucheng@sii.edu.cn\\
zxzyin@cee.ecnu.edu.cn
}

\begin{document}

\maketitle

\begin{abstract}
With the rapid development of AIGC technologies, generative image steganography has attracted increasing attention due to its high imperceptibility and flexibility. However, existing generative steganography methods often maintain acceptable security and robustness only at relatively low embedding rates, severely limiting the practical applicability of steganographic systems. To address this issue, we propose a novel DTAMS framework that achieves high embedding rates while ensuring strong robustness and security. Specifically, a dynamic multi-timestep adaptive embedding mechanism is constructed based on transition-cost modeling in diffusion models, enabling automatic selection of optimal embedding timesteps to improve embedding rates while preserving overall performance. Meanwhile, we propose a global sub-interval mapping strategy that jointly considers mapping errors and the frequency distribution of secret information, converting point-wise perturbations into interval-level statistical mappings to suppress error accumulation and distribution drift during multi-step diffusion processes. Furthermore, a multi-dimensional joint constraint mechanism is introduced to mitigate distortions caused by repeated latent-pixel transformations by jointly regularizing embedding errors at the pixel, latent, and semantic levels. Experiments demonstrate that the proposed method achieves an embedding rate of 12 bpp while maintaining excellent security and robustness. Across all evaluated conditions, DTAMS reduces the average extraction error rate by 59.39\%, representing a significant improvement over SOTA methods.
\end{abstract}

\section{Introduction}

With the rapid advancement of AIGC technology, concerns regarding data security and privacy protection have become increasingly prominent. Steganography, as a core information hiding \cite{gao2026survey,guan2025non,bao2024pluggable,wu2024generative} technique, enables imperceptible transmission of secret information by embedding information into cover such as text \cite{zhou2025auto,long2025scf}, audio \cite{li2025coas,Yan2025FGSAudioFF}, image \cite{qi2025provably,cheng2025robust}, or video \cite{Meng2025Video,mao2024covert}. Recent progress in generative modeling has enabled generative steganography, integrating secret embedding into the cover generation process to enable unified synthesis and hiding with enhanced imperceptibility and flexible embedding.

\begin{figure}[t]
    \centering
    \includegraphics[width=0.7\linewidth]{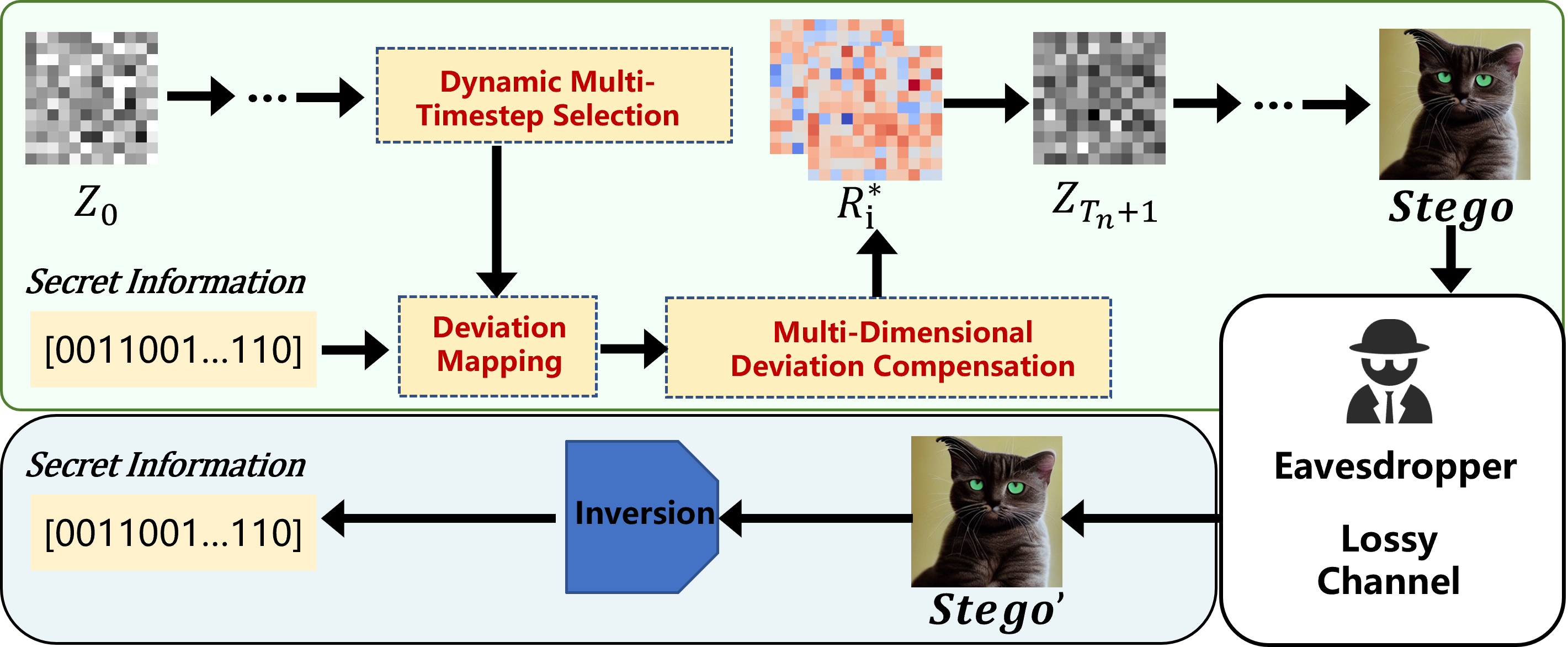}  
    \caption{The pipeline of DTMAS. }
    \label{fig:first}
\end{figure}
Existing generative image steganography methods \cite{yang2024diffstega,luo2024comprehensive,ma2023robust} can be broadly categorized by the underlying generative models they use. Early GAN-based approaches, such as DCGAN-Steg \cite{Hu2018DCGANStego}, produce visually plausible images but are often constrained by model selection, leading to limited fidelity and unstable training. IDEAS \cite{Liu2022DisentanglementAE} improves decoding reliability by embedding secrets into disentangled structural latent representations, while flow-based methods such as S2IRT \cite{Zhou2022SecretToImage} leverage invertible latent--image mappings to enable deterministic embedding and extraction, albeit at the cost of increased computational complexity. Despite recent progress, existing methods are vulnerable to quantization and lack controllable generation tied to sender attributes or covert patterns.

More recently, diffusion probabilistic models and their latent-space variants have emerged as a mainstream paradigm for generative image steganography, owing to their stable training dynamics and high-quality image synthesis. Leveraging these properties, existing diffusion-based methods integrate secret embedding into diffusion image generation. StegaDDPM \cite{Peng2023Stegaddpm} embeds secret information into diffusion residuals at intermediate timesteps, but its dependence on fragile statistical properties limits robustness as embedding capacity increases. LDStega \cite{Peng2024LDStega} performs embedding in the latent residual space of latent diffusion models via truncated Gaussian encoding, where strong latent constraints improve decoding stability but severely restrict embedding capacity. MDDM \cite{xu2025mddm} embeds secret bits at the initial noise stage via message-driven noise construction, thereby reducing controllability and leading to robustness degradation during denoising. DGADM-GIS \cite{Yuan2025DGADM} adopts deterministic guided additive sampling at the same early diffusion stage and similarly suffers from deviation amplification under common image distortions. Overall, existing diffusion-based steganography approaches remain insufficiently robust in high-capacity regimes, constraining their applicability.

To mitigate the limitations described above, we propose DTAMS, a diffusion-based generative steganography framework that integrates dynamic multi-timestep selection with adaptive deviation mapping. As illustrated in Fig.~\ref{fig:first}, rather than focusing on low-payload concealment performance, DTAMS is designed to systematically expand the exploitable embedding space within diffusion models by identifying and leveraging diffusion timesteps that are more robust to embedding-induced statistical deviations. Within this framework, a dynamic multi-timestep adaptive embedding mechanism is introduced to embed secret information into diffusion stages that exhibit stronger robustness and security. Meanwhile, a global sub-interval deviation mapping strategy reformulates the embedding process as a statistically constrained interval-level transformation, enabling a large embedding rate. In addition, a lightweight multi-dimensional regulation scheme is incorporated to constrain cumulative deviations introduced by repeated latent-pixel transformations. Together, these components enable substantially increased embedding rate while ensuring robustness and anti-steganalysis performance. Experimental results demonstrate that the proposed method achieves an embedding capacity of 12 bpp with an information extraction accuracy exceeding 99.2\%, while preserving high perceptual quality with an average PSNR of 33.24 dB and SSIM of 0.9865. The main contributions of this work are summarized as follows:

\begin{itemize}
\item \textbf{A dynamic multi-timestep adaptive embedding mechanism precomputes transformation costs at different timesteps and selects optimal timesteps for information embedding, thereby achieving a high embedding rate while ensuring robustness and security.}

\item \textbf{A global sub-interval deviation mapping strategy uses interval-level transformations, refined with linear residual integration and full-image optimization, suppressing deviation accumulation under high embedding rates.}

\item \textbf{A multi-dimensional deviation compensation mechanism jointly constrains pixel, latent, and semantic deviations, mitigating distortions from repeated latent-pixel transformations and enhancing robustness.}

\end{itemize}

\section{Related Work}

\subsection{Diffusion Models}

Diffusion models have emerged as a prominent class of generative models due to their stable training dynamics and capability to produce high-fidelity images. Representative works, such as Denoising Diffusion Probabilistic Models (DDPM) \cite{ho2020denoising}, demonstrate that iterative denoising can yield high-quality images comparable to those of GAN-based approaches. Subsequent developments, including Denoising Diffusion Implicit Models (DDIM) \cite{song2021denoising}  and latent diffusion models (LDMs) \cite{rombach2022high,Lugmayr2022repaint}, improve sampling efficiency through deterministic updates or by performing diffusion in a lower-dimensional latent space. In addition, LDMs leverage pre-trained autoencoders to encode images into compact latent representations, allowing for controllable conditional generation with semantic consistency. Due to their inherent iterative and controllable generation process, diffusion models have recently become attractive for generative steganography, providing a structured and robust framework for embedding secret information while maintaining stego image quality.

\subsection{Generative Image Steganography}

Generative image steganography (GIS) \cite{Cheng2025RFNNs} represents a paradigm shift from traditional embedding-based methods. Instead of modifying existing cover images, GIS hides secret information during image synthesis, producing stego images that naturally resemble ordinary images. This approach avoids introducing detectable artifacts, thus improving security against advanced steganalysis.

Early GIS was largely GAN-based: DCGAN-Stega \cite{Hu2018DCGANStego} encoded secrets in latent noise to synthesize stego images, but often faced a three-way trade-off among fidelity, capacity, and controllability. Later work improved stability and recovery. IDEAS \cite{Liu2022DisentanglementAE} disentangled structure and texture in an autoencoder to support more reliable embedding, while S2IRT \cite{Zhou2022SecretToImage} leveraged Glow-style invertible mappings for reversible hiding with accurate extraction. Diffusion models then expanded the design space. StegaDDPM \cite{Peng2023Stegaddpm} embedded high-capacity data through DDPM intermediate states while keeping outputs visually natural, and LDStega \cite{Peng2024LDStega} paired latent diffusion with truncated Gaussian sampling for better control, robustness, and format compatibility. In parallel, CRoSS \cite{Yu2023CROSS} explored image-to-image pipelines, trading off flexibility for an alternative embedding route.



Despite these advances, current GIS approaches still face challenges. High embedding rate often leads to cumulative distortions in the generated images, whereas real-world transmission conditions, such as compression or noise, can degrade secret recovery.

\begin{figure*}[htbp]
    \centering
    \includegraphics[width=0.7\linewidth]{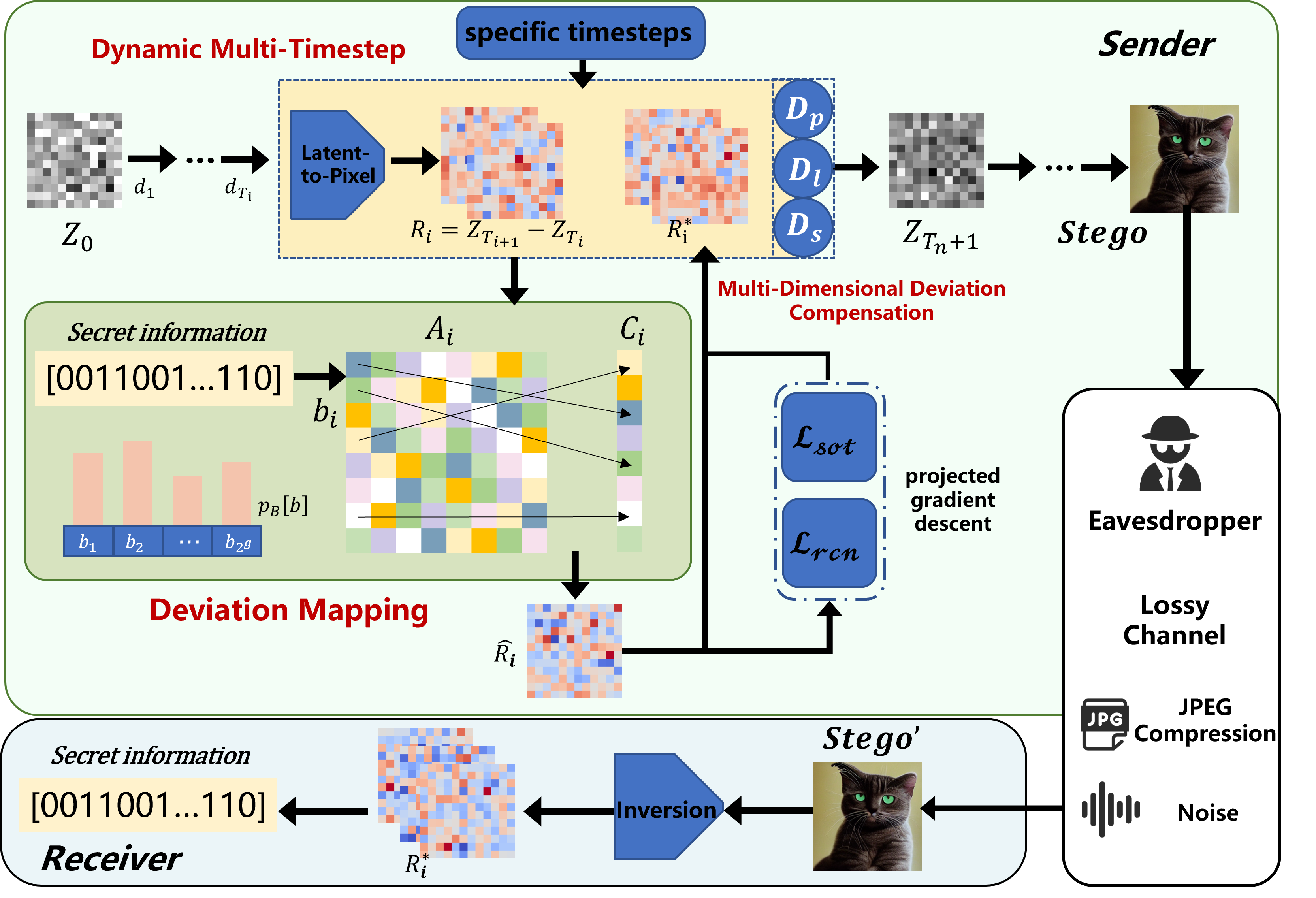}
    \caption{The Framework of DTAMS. }
    \label{fig:pipeline}
\end{figure*}

\section{Methodology}

\subsection{Framework Overview and Processing Pipeline}

The framework Dynamic Multi-Timestep Selection and Adaptive Deviation Mapping for Diffusion Steganography (DTAMS) is proposed to address the challenge of achieving high embedding rates while ensuring robust and anti-steganalysis performance.

Taking  Fig.~\ref{fig:pipeline} as an illustration, on the sender side, DTAMS first generates the intermediate states of the diffusion process, along with the unembedded cover content, from an initial latent-space seed. A dynamic multi-timestep adaptive embedding mechanism is then employed to evaluate the transformation cost of intermediate states at different diffusion timesteps and to select an optimal subset for embedding secret information. For each selected intermediate state, DTAMS embeds the secret information by applying a global sub-interval mapping strategy that maps pixel-value intervals to corresponding target intervals. After interval mapping, a full-image interval optimization based on projected gradient descent is performed to refine the mapped image and suppress high-frequency noise, producing the final embedding residual that is fed back into the diffusion process. During the transformation from pixel space back to latent space, DTAMS introduces a multi-dimensional joint-constraint mechanism to regularize embedding-induced deviations across the pixel, latent, and semantic spaces. The optimized residuals are subsequently integrated into the diffusion pipeline, resulting in the final stego image.

On the receiver side, DTAMS reconstructs the corresponding intermediate states, extracts residuals at the shared embedding timesteps, and recovers the secret information via the shared mapping matrices.

\begin{table}[t]
    \centering
    \begin{small}
    \begin{tabular}{ll}
    \toprule
    \textbf{Notation} & \textbf{Description} \\
    \midrule
    $T$          & Set of diffusion timesteps \\
    $T^*$        & Selected embedding timesteps \\
    $A_i$        & Original quantization interval \\
    $C_j$        & Target embedding interval \\
    $P^*$        & Globally optimal interval mapping \\
    $\mathrm{cost}_{AC}$ 
                 & Mapping cost from interval $A$ to $C$ \\
    $p_B[b]$     & Probability of secret symbol $b$ \\
    $N(b)$       & Occurrence count of symbol $b$ \\
    $\mathscr{L}_{rcn}$    & Reconstruction loss \\
    $\mathscr{L}_{sot}$    & Smoothness loss \\
    $w$          & Fixed weighting coefficient for response fusion \\
    $\alpha_p,\alpha_l,\alpha_s$ 
                 & Balancing coefficients for multi-level losses \\
    \bottomrule
    \end{tabular}
    \end{small}
    \caption{Notations.}
    \label{tab:notation}
\end{table}

\subsection{Dynamic Multi-Timestep Adaptive Embedding via Preprocessing-Based Selection}

Previous studies have shown that intermediate states at different diffusion timesteps exhibit heterogeneous tolerance to embedding-induced deviations and varying error-attenuation behaviors during subsequent denoising steps. Motivated by this, we propose a dynamic multi-timestep adaptive embedding mechanism, which precomputes the embedding-induced transformation cost at each timestep and selects an optimal subset that minimizes cumulative distortion while maintaining a balance between security and robustness.

The generative process of diffusion models can be formally modeled as a Markov chain, where the state at each timestep depends only on the previous state. Specifically, the transition from an intermediate state $X_t$ to the preceding state $X_{t-1}$ can be expressed as:
\begin{equation}
X_{t-1} = \mu_t + \sigma_t \cdot Z_t,
\end{equation}
where $Z_t \sim \mathcal{N}(0, I)$ denotes Gaussian noise, $\mu_t$ is the model-predicted mean, and $\sigma_t$ is the predefined variance schedule. Based on this formulation, the embedding-induced transformation cost can be precomputed for each intermediate state, providing a principled criterion for selecting a subset of timesteps that minimizes cumulative distortion while maintaining a balance between security and robustness.

\begin{figure}[b]
    \centering
    \includegraphics[width=0.7\linewidth]{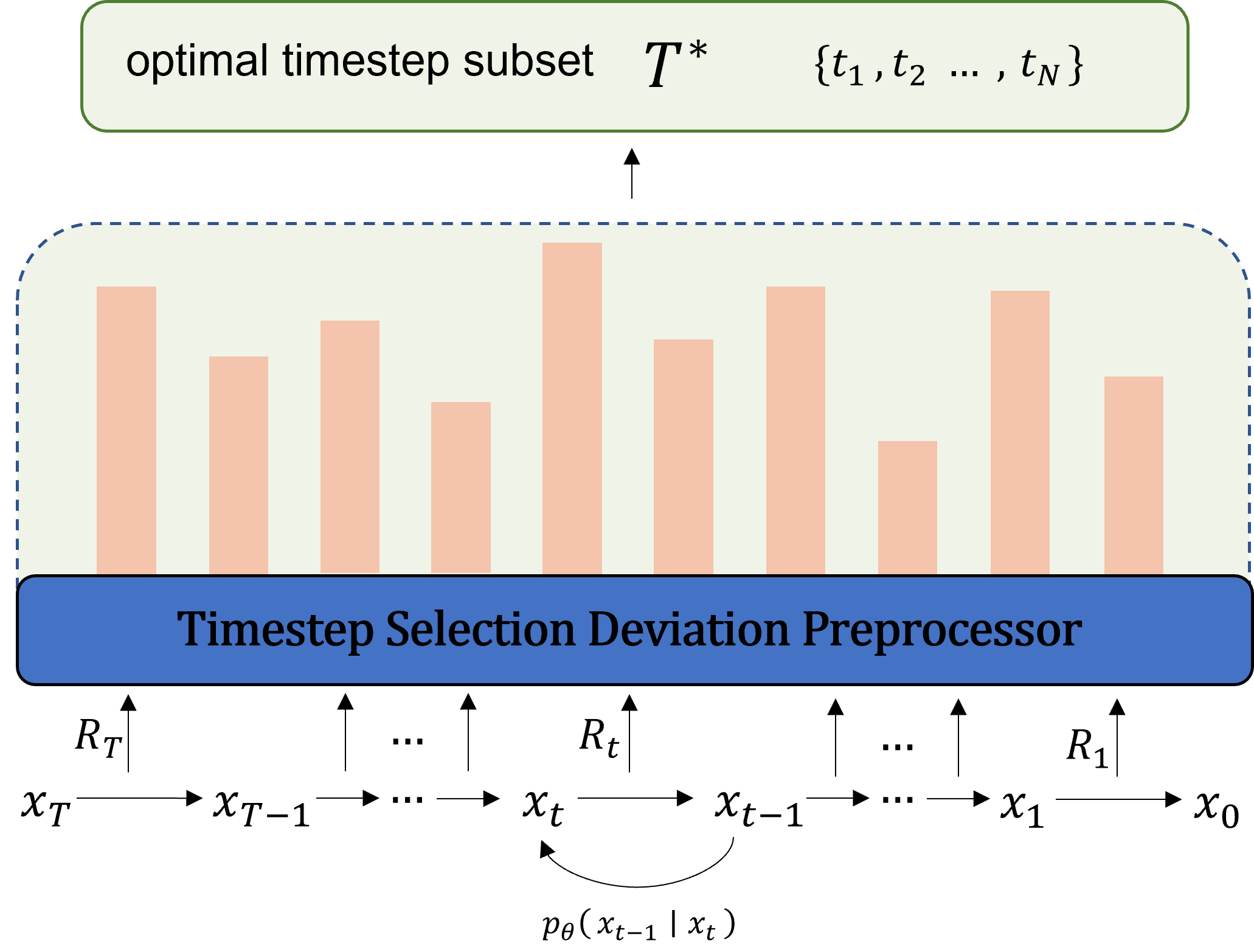}
    \caption{Multi-Timestep Adaptive Embedding and Selection}
    \label{fig:timestep}
\end{figure}

The weighted transformation cost is computed in advance for each timestep, and an optimal subset of timesteps $T^*$ is selected to minimize the overall embedding-induced distortion. Formally, as illustrated in Fig.~\ref{fig:timestep}, the optimal timestep subset is determined as
\begin{equation}
T_i^* = \arg\min_{T'\subseteq T,\,|T'|=n}\sum_{t\in T'} \overline{\mathrm{cost}}_{AC}[A][C],\quad i=1,\ldots,k.
\end{equation}

In practice, timesteps from the middle-to-late diffusion stages are preferred, as high-frequency details generated at these stages render embedding modifications less perceptible. Additionally, subsequent denoising steps mitigate noise accumulation and quantization errors, enhancing embedding stability and overall robustness. Leveraging these strategies, the proposed mechanism effectively addresses the inherent trade-off between security and robustness that constrains conventional fixed-timestep approaches.

\subsection{Sub-Interval Partitioning Based on Adaptive Deviation Mapping}

Although optimal timestep selection reduces global deviations, direct modifications of intermediate states based on secret information still lead to local deviation accumulation across multiple diffusion steps. These deviations propagate and amplify during iterative denoising, causing shifts in pixel-value distributions and increased decoding errors, especially under high-capacity embedding or images with non-uniform pixel distributions. To mitigate this, a global sub-interval mapping strategy is applied to the selected embedding timesteps. Instead of modifying individual pixels, the pixel-value intervals of the original intermediate states are mapped to corresponding intervals in the embedded states according to the secret information. Operating at the interval level constrains local modifications within statistically consistent ranges, thereby suppressing deviation propagation and dynamic distortion while maintaining the overall pixel distribution, which preserves both the statistical consistency and steganographic security of the stego.

For a given intermediate state $Z_t$, we estimate the global pixel-value distribution and partition the value range into $T=2^g$ sub-intervals using the inverse cumulative distribution function $F^{-1}(\cdot)$. This quantile-based partitioning ensures that each sub-interval contains approximately the same proportion of pixels, thereby preventing large fluctuations in deviation due to uneven interval sizes. For any original interval $A$ and target interval $C$, we define the base mapping cost as the expected mean squared deviation.
\begin{equation}
\mathrm{cost}_{AC}=\mathbb{E}_{\mu\sim A}\left[(\mu-c_C)^2\right],
\end{equation}
where $c_C$ is typically the midpoint of interval $C$.

\begin{algorithm}[t]
\caption{Global Optimal Sub-Interval Mapping}
\label{alg:global_interval_mapping}
\begin{algorithmic}[1]
\Require Intermediate state $Z_t$, number of intervals $T = 2^g$, inverse CDF $F^{-1}(\cdot)$, symbol counts $\{N(b)\}_{b \in \mathscr{B}}$
\Ensure Quantile intervals $\{A_i\}_{i=1}^T$, target intervals $\{C_j\}_{j=1}^T$, optimal mapping $P^*$

\State $M \gets \sum_{b \in \mathscr{B}} N(b)$

\For{$i = 0$ to $T$}
    \State $\tau_i \gets F^{-1}(i/T)$
\EndFor

\For{$i = 1$ to $T$}
    \State $A_i \gets [\tau_{i-1}, \tau_i)$
    \State $C_i \gets A_i$
    \State $b_i \gets$ symbol associated with $A_i$
    \State $p_i \gets N(b_i)/M$
\EndFor

\For{$i = 1$ to $T$}
    \For{$j = 1$ to $T$}
        \State $\mathrm{cost}_{ij} \gets \mathbb{E}_{\mu \sim A_i}\big[(\mu - c_{C_j})^2\big]$
        \State $\overline{\mathrm{cost}}_{ij} \gets p_i \cdot \mathrm{cost}_{ij}$
    \EndFor
\EndFor

\State $P^* \gets \arg\min_{P \in \mathfrak{S}_T} \sum_{i=1}^{T} \overline{\mathrm{cost}}_{i,P(i)}$

\State \Return $\{A_i\}_{i=1}^T,\ \{C_j\}_{j=1}^T,\ P^*$

\end{algorithmic}
\end{algorithm}

To account for the non-uniform distribution of secret symbols, we incorporate the symbol probability $p_B[b]$ into the cost model. Let
\begin{equation}
p_B[b]=\frac{N(b)}{\sum_k N(b_k)},
\end{equation}
where $N(b)$ denotes the occurrence count of symbol $b$. The resulting weighted global mapping cost is
\begin{equation}
\overline{\mathrm{cost}}_{AC}[A][C]=p_B[b]\cdot \mathrm{cost}_{AC}[A][C].
\end{equation}

Overall, our objective is to construct a bijective symbol-to-interval mapping that minimizes the total weighted distortion:
\begin{equation}
P^* = \arg\min_{P\in\Pi}\sum_{b\in B}\overline{\mathrm{cost}}_{AC}\left[A_b\right]\left[C_{P(b)}\right].
\end{equation}
Solving the resulting cost matrix yields the globally optimal discrete interval mapping, which reduces distribution deviations induced by embedding. The detailed procedure for computing the total weighted distortion $P^*$ using DTAMS is outlined in Algorithm~\ref{alg:global_interval_mapping}.

\begin{figure}[b]
    \centering
    \includegraphics[width=0.66\linewidth]{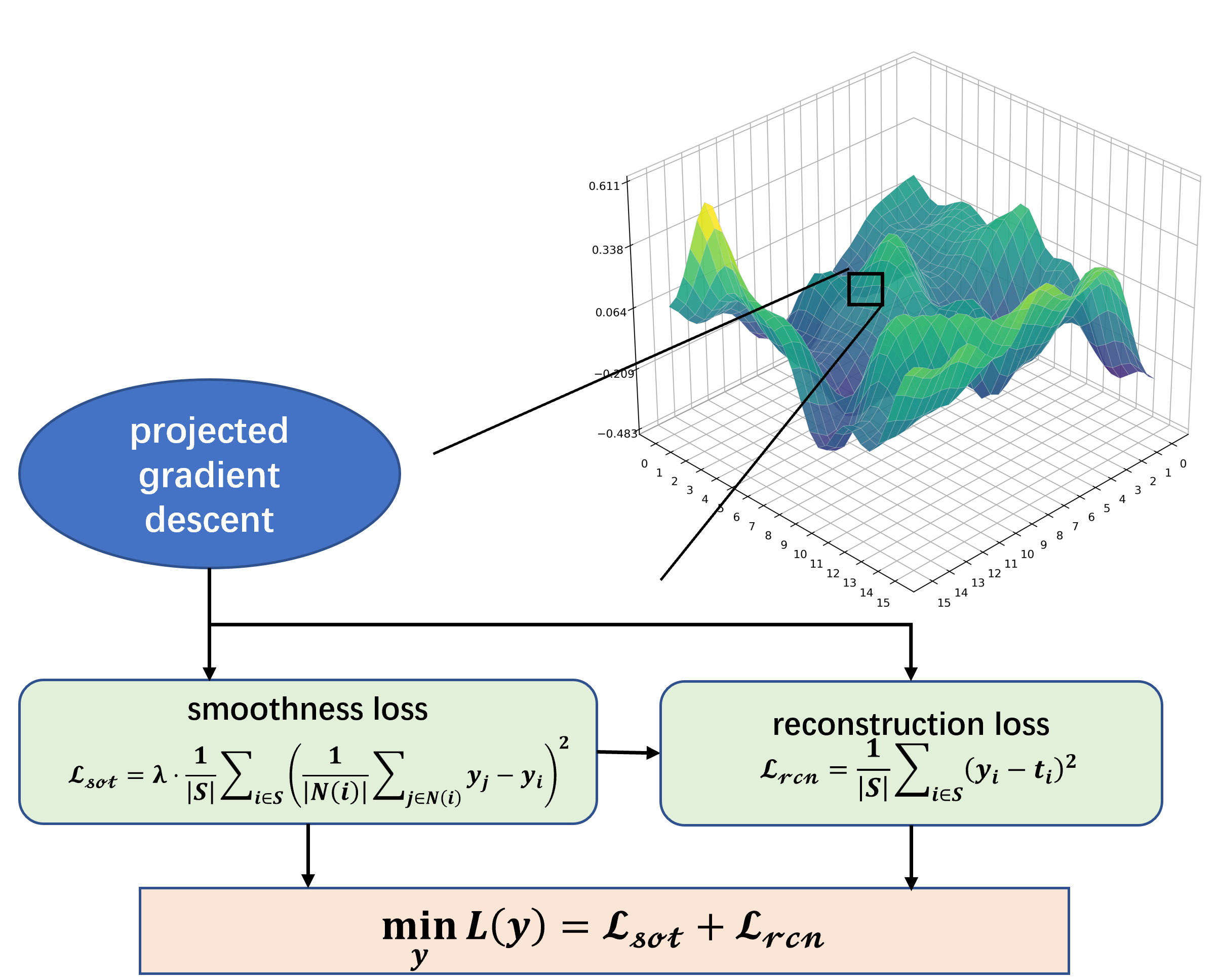}
    \caption{The PGD full-image interval optimization highlights the combined effect of smoothness and reconstruction losses.}
    \label{fig:pgd}
\end{figure}

After constructing the globally optimal discrete interval mapping, we further refine the embedded image via full-image interval optimization using projected gradient descent. This optimization mitigates high-frequency noise introduced during embedding and enhances the robustness of the hidden secret. As illustrated in Fig.~\ref{fig:pgd} , the optimization is guided by two complementary loss functions: a smoothness loss that enforces local pixel consistency,
\begin{equation}
\mathscr{L}_{sot} = \lambda \cdot \frac{1}{|S|} \sum_{i \in S} \left( \frac{1}{|N(i)|} \sum_{j \in N(i)} y_j - y_i \right)^2,
\end{equation}
and a reconstruction loss that constrains the optimized pixels to remain close to the original interval values,
\begin{equation}
\mathscr{L}_{rcn} = \frac{1}{|S|} \sum_{i \in S} (y_i - t_i)^2.
\end{equation}
The overall optimization problem is formulated as
\begin{equation}
\min_y \left( \mathscr{L}_{rcn} + \mathscr{L}_{sot} \right), \quad y_i \in [t_i - \epsilon, t_i + \epsilon],
\end{equation}
where \(y_i\) denotes the optimized pixel value, \(t_i\) the target interval value, and \(N(i)\) the neighborhood of pixel \(i\). By applying this full-image optimization, the intermediate residuals are adjusted to be statistically imperceptible, while preserving both global and local statistics, thereby achieving a high level of steganographic undetectability.

\begin{table*}[t]
\centering
\begin{tabular}{cccccc}
\toprule
timesteps & ER (bpp) & MAE ($\times 10^{-2}$) & PSNR (dB) & SSIM & LPIPS ($\times 10^{-2}$) \\
\midrule
1 & 2.4 & \textbf{0.4228} & \textbf{42.3490} & \textbf{0.9977} & \textbf{0.0819} \\
2 & 4.8 & 0.5945 & 39.7776 & 0.9965 & 0.1345 \\
3 & 7.2 & 0.8235 & 36.9570 & 0.9937 & 0.2382 \\
4 & 9.8 & 1.0809 & 34.6139 & 0.9896 & 0.3722 \\
5 & 12 & 1.2968 & 33.0151 & 0.9864 & 0.5097 \\
6 & 14.4 & 1.5202 & 31.7367 & 0.9812 & 0.7316 \\
7 & 16.8 & 1.7387 & 30.6583 & 0.9774 & 0.8712 \\
8 & \textbf{19.6} & 1.9663 & 29.4370 & 0.9721 & 1.0588 \\
\bottomrule
\end{tabular}
\caption{Image quality under different timesteps.}
\label{tab:reconstruction_metrics}
\end{table*}

After the full-image interval optimization, the final embedded residual is obtained by linearly combining the optimized intermediate state with the original intermediate state. Formally, let $Z_t$ denote the original intermediate state at timestep $t$, and $\hat{Z}_t$ the optimized embedded state obtained after the interval mapping and gradient-based refinement. The final residual $R_t$ is computed as

\begin{equation}
R_t = w\,\hat{Z}_t + (1-w)\,Z_t,
\label{eq:weighted_rt}
\end{equation}
where $w \in [0,1]$ is a balancing coefficient that controls the relative contribution.

Consequently, the embedded pixel-value distribution remains close to the original statistics, enabling full-image high-capacity embedding while maintaining strong steganographic security.

\begin{table}[htbp]
  \centering
  \setlength{\tabcolsep}{3pt}
  \begin{tabular}{lccccc}
    \toprule
    \multirow{2}{*}{Approaches} 
      & \multicolumn{3}{c}{Acc (\%)} 
      & \multirow{2}{*}{ER (bpp)} 
      & \multirow{2}{*}{Image size} \\
    \cmidrule(lr){2-4}
      & FFHQ & Bedroom & Cat & & \\
    \midrule
    IDEAS      & 61.83 & 60.57 & 61.89 & 3.125e-2 & 256$\times$256 \\
    S2IRT      & 71.73 & 71.05 & 71.46 & 3        & 64$\times$64 \\
    StegaDDPM  & 96.81 & 94.25 & 95.34 & 9        & 256$\times$256 \\
    LDStega    & 99.22 & 99.36 & 99.28 & 4.28e-2  & 256$\times$256 \\
    \textbf{Ours} 
               & \textbf{99.77} & \textbf{99.99} & \textbf{99.41} & \textbf{12} & 256$\times$256 \\
    \bottomrule
  \end{tabular}
  \caption{Comparison of extraction accuracy without attacks.}
  \label{tab:booktabs}
\end{table}

\subsection{Multi-Dimensional Deviation Compensation}

In diffusion-based generative steganography, intermediate states undergo repeated transformations between the latent and pixel spaces, introducing deviations such as quantization noise, latent distribution drift, and semantic inconsistencies. Reliance on a single constraint is typically insufficient to effectively mitigate these deviations. To address this, a multidimensional error-compensation mechanism is introduced. Pixel-level constraints preserve low-level structures, latent-space regularization maintains the statistical properties of latent representations, and semantic constraints reduce high-level inconsistencies. By integrating these constraints into a unified optimization objective, the mechanism effectively suppresses the accumulation of deviations caused by repeated latent-pixel transformations, thereby enhancing both image quality and steganalysis resistance.

This multi-dimensional, cross-space design improves embedding stability and message recoverability under common image distortions. By jointly modeling deviations along pixel, latent, and semantic dimensions, the mechanism provides moderate robustness enhancement without introducing significant computational overhead.

To compensate for deviations introduced by repeated latent-pixel transformations, DTAMS introduces a multidimensional error-compensation mechanism that jointly constrains distortions across different representation levels. Specifically, pixel-level deviation $D_p$ preserves low-level structural fidelity between the stego state and the original image, latent-space deviation $D_l$ regularizes embedded latent variables to remain consistent with the diffusion prior distribution, and semantic deviation $D_s$ enforces high-level perceptual consistency to suppress semantic drift during generation. By jointly modeling these complementary deviations within a unified optimization objective, DTAMS effectively mitigates cumulative transformation deviations while improving visual quality and robustness against common image distortions.

The overall compensation objective is formulated as a weighted combination of the above losses:
\begin{equation}
L_{\text{cm}} = \alpha_p D_p + \alpha_l D_l + \alpha_s D_s,
\end{equation}

where $\alpha_p$, $\alpha_l$, and $\alpha_s$ are pre-defined balancing coefficients. Optimization is performed via backpropagation to simultaneously mitigate transformation distortion and preserve image characteristics.

\begin{table*}[t]
\centering
\setlength{\tabcolsep}{4pt}
\renewcommand{\arraystretch}{0.9}
\begin{tabular}{llccccccccc}
\toprule
\multirow{3}{*}{Distortion Type} & \multirow{3}{*}{Ratio}
& \multicolumn{9}{c}{Acc (\%)} \\
\cmidrule(lr){3-11}
& & \multicolumn{3}{c}{StegaDDPM} & \multicolumn{3}{c}{LDStega} & \multicolumn{3}{c}{Ours} \\
\cmidrule(lr){3-5} \cmidrule(lr){6-8} \cmidrule(lr){9-11}
& & Cat & FFHQ & Bedrooms & Cat & FFHQ & Bedrooms & Cat & FFHQ & Bedrooms \\
\midrule
\multirow{3}{*}{Gaussian noise}
& 0.07\% &70.95 & 66.83 & 67.54 & 95.18 & 92.68 & 91.24 & \textbf{98.41} & \textbf{98.52} & \textbf{97.63} \\
& 0.04\% &73.18 & 68.24 & 69.05 & 96.03 & 97.24 & 94.87 & \textbf{98.78} & \textbf{98.82} & \textbf{98.05} \\
& 0.01\% &83.45 & 77.62 & 78.31 & 99.07 & 98.19 & 98.35 & \textbf{99.42} & \textbf{99.71} & \textbf{99.66} \\
\midrule
\multirow{3}{*}{Salt \& Pepper noise}
& 0.07\% & 94.60 & 93.45 & 92.42 & 94.49 & 93.91 & 93.31 & \textbf{98.03} & \textbf{98.14} & \textbf{98.21} \\
& 0.04\% & 93.98 & 91.29 & 90.91 & 96.41 & 95.94 & 95.66 & \textbf{98.59} & \textbf{98.62} & \textbf{98.67} \\
& 0.01\% & 93.32 & 90.18 & 90.39 & 99.02 & 98.31 & 98.64 & \textbf{99.31} & \textbf{99.22} & \textbf{99.34} \\
\midrule
\multirow{2}{*}{JPEG}
& 90\% &63.74 & 61.88 & 61.52 & 99.48 & 98.19 & \textbf{99.49} & \textbf{99.68} & \textbf{99.12} & 98.24 \\
& 70\% & 59.10 & 59.21 & 58.86 & \textbf{98.36} & 97.46 & 96.91 & 98.33 & \textbf{98.71} & \textbf{97.58} \\
\bottomrule
\end{tabular}
\caption{Robustness comparison under different image distortions.}
\label{tab:robustness}
\end{table*}

\begin{table*}[t]
\centering
\begin{tabular}{l l c c c c c}
\toprule
Dataset & Steganalysis & IDEAS & S2IRT & StegaDDPM & LDStega & Ours \\
\midrule
\multirow{3}{*}{FFHQ} 
& YeNet & 0.5203 & 1.0000 & 0.5108 & 0.9965 & \textbf{0.5097} \\
& SRNet & 0.5095 & 0.9902 & 0.5003 & 0.9957 & \textbf{0.4998} \\
& SiaStegNet & 0.5102 & 0.9898 & 0.4995 & 0.9979 & \textbf{0.5001} \\
\midrule
\multirow{3}{*}{Bedroom} 
& YeNet & 0.5302 & 0.9901 & 0.5005 & 0.9873 & \textbf{0.4999} \\
& SRNet & 0.5198 & 0.9805 & 0.4943 & 0.9967 & \textbf{0.5012} \\
& SiaStegNet & 0.5101 & 0.9973 & \textbf{0.4978} & 0.9983 & 0.5063 \\
\midrule
\multirow{3}{*}{Cat} 
& YeNet & 0.5008 & 0.9902 & 0.5011 & 0.9845 & \textbf{0.4998} \\
& SRNet & 0.5201 & 0.9899 & \textbf{0.4999} & 0.9698 & 0.5002 \\
& SiaStegNet & 0.5198 & 1.0000 & 0.5112 & 0.9968 & \textbf{0.5004} \\
\bottomrule
\end{tabular}
\caption{The comparison of the anti-steganalysis performance under different steganalysis networks.
}
\label{tab:steganalysis}
\end{table*}

\begin{table}[t]
\centering
\begin{tabular}{c c c}
\toprule
Method & Embedding Rate (bpp) & MAE ($\times 10^{-2}$) \\
\midrule
\multirow{2}{*}{StegaDDPM} & 3 & 0.6427 \\
                           & 9 & 0.6789 \\
\midrule
LDStega & $4.28 \times 10^{-2}$ & 5.8957 \\
\midrule
\multirow{2}{*}{Ours} & 4.8 & \textbf{0.5763} \\
                      & \textbf{12} & 1.2774 \\
\bottomrule
\end{tabular}
\caption{Comparison of embedding rate and MAE.}
\label{tab:embedding_comparison}
\end{table}

\section{Experiments}

This section presents the experimental setup and results. Section 4.1 describes the setup. Sections 4.2 and 4.3 report capacity flexibility and practicality. Sections 4.4 and 4.5 assess the robustness and security of DTAMS.

\subsection{Experimental Setup}

In our experiments, we employed a publicly pretrained Latent Diffusion Model to perform generative image steganography. The latent space has dimensions of $32 \times 32 \times 4$, with a sub-interval granularity $g=3$ and a balancing coefficient $w=0.75$. Both forward and backward diffusion processes consisted of 100 steps. All experiments were conducted on a vGPU with 32 GB memory and a 16-core Xeon(R) Platinum 8352V CPU. To regulate multi-dimensional embedding deviations, the compensation losses were weighted as $\alpha_p = 1.0$ for pixel-level constraints, $\alpha_s = 0.75$ for semantic-level constraints, and $\alpha_l = 0.005$ for latent-space constraints. We evaluated the proposed DTAMS method against four state-of-the-art generative steganography methods, namely IDEAS \cite{Liu2022DisentanglementAE}, S2IRT \cite{Zhou2022SecretToImage}, StegaDDPM \cite{Peng2023Stegaddpm}, and LDStega \cite{Peng2024LDStega}, using the Bedroom and Cat datasets \cite{Yu2015LSUN}, as well as face images from FFHQ \cite{Karras2019StyleGAN}.

\subsection{Flexibility of Steganographic Capacity}

To validate the capacity flexibility of DTAMS, the number of selected embedding timesteps is varied from 1 to 8, thereby controlling the total embedding payload. Table~\ref{tab:reconstruction_metrics} reports the steganographic capacity and information extraction accuracy corresponding to different numbers of embedding timesteps. The results indicate that secret information extraction accuracy remains consistently high across all capacity levels, demonstrating that DTAMS preserves reliable message recovery even as the embedding payload increases. Furthermore, image quality metrics (PSNR and SSIM) show minimal degradation with increased capacity, confirming that the timestep selection strategy effectively balances embedding capacity with perceptual fidelity.

\begin{figure}[htbp]
    \centering
    \includegraphics[width=0.85\linewidth,height=0.4\linewidth]{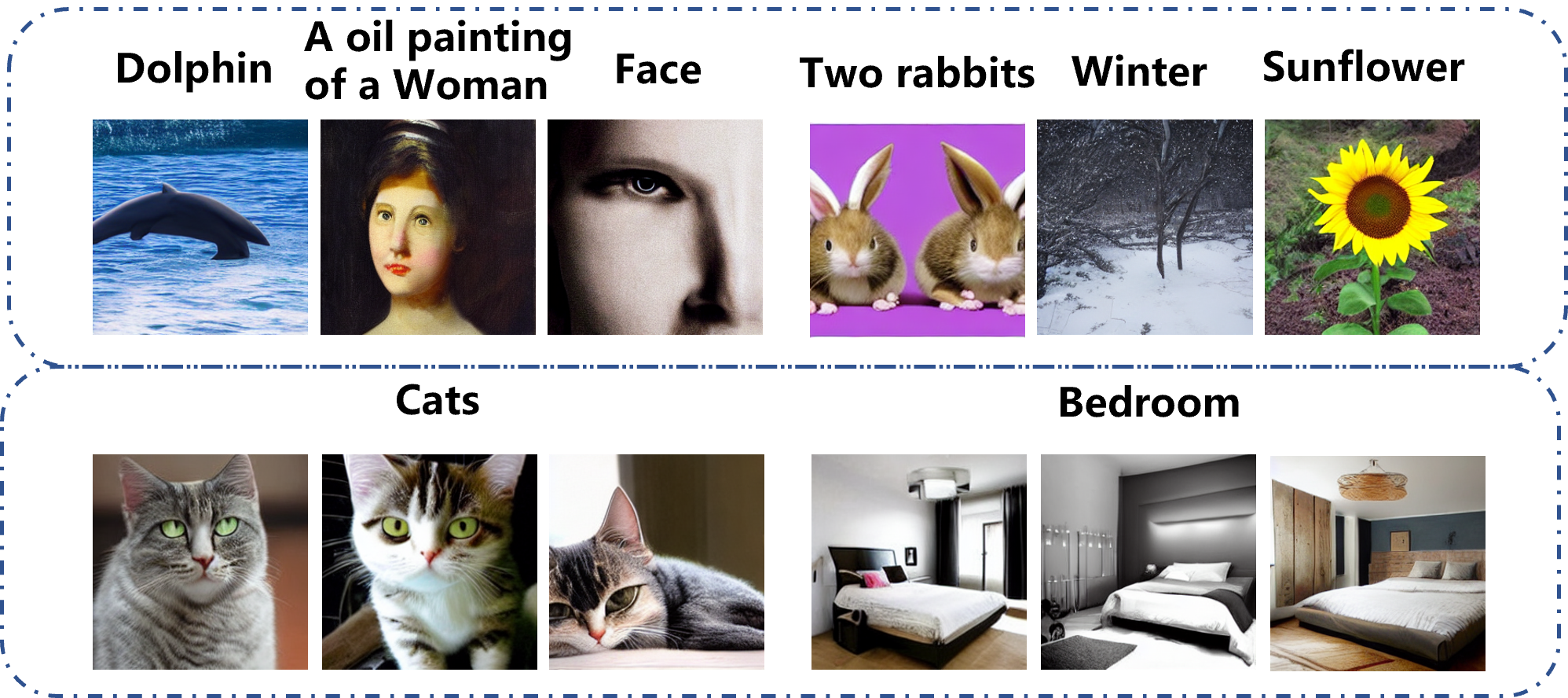}
    \caption{DTAMS generates stego images conditioned on input text, producing personalized images for different scenes and multiple variations for the same scene.}
    \label{fig:photos}
\end{figure}

\subsection{Practicality}
In image steganography, practicality is typically evaluated in terms of controllability and availability.

\noindent\textbf{Controllability.} We evaluate the controllability of DTAMS in generating stego images with specified content conditions. Figure~\ref{fig:photos} illustrates two scenarios. In the first scenario, distinct input conditions generate personalized stego images tailored to the described scene. In the second scenario, the same input condition produces multiple stego images with consistent semantic characteristics. DTAMS exhibits precise control over generated content, including object type, quantity, and style, surpassing existing GIS methods such as IDEAS, S2IRT, and StegaDDPM, which are limited to the distribution of training data. This controllability ensures that the presence of embedded information cannot be inferred from the generated content, enhancing the security of the steganographic framework.

\noindent\textbf{Availability.} To assess practical availability, stego images generated by DTAMS are discretized and saved in PNG and JPEG formats. Table~\ref{tab:booktabs} reports embedding rate, extraction accuracy, and image size. DTAMS maintains high extraction accuracy across all datasets and formats, demonstrating strong resistance to quantization errors and compression artifacts. In comparison, IDEAS and S2IRT exhibit lower extraction accuracy under these conditions, indicating reduced robustness to common image processing operations.

\subsection{Robustness}
\noindent\textbf{Robustness under non-attack conditions.} Table~\ref{tab:booktabs} reports the secret extraction accuracy of DTAMS under non-attack conditions. Across all datasets, DTAMS consistently maintains high accuracy, with PSNR and SSIM values remaining close to those of the original image. Even at high embedding rates, the method maintains satisfactory visual fidelity and reliably recovers hidden information.

\noindent\textbf{Robustness with attack conditions.} Stego images transmitted through communication channels inevitably encounter various perturbations, such as Gaussian noise, salt-and-pepper noise, and JPEG compression. These disturbances can degrade extraction accuracy, compromising the reliability of covert communication. As shown in Table~\ref{tab:robustness}, DTAMS consistently preserves accuracy under these conditions. Notably, the average error rate across all perturbations is reduced by 59.39\% compared with competing methods. This demonstrates that the proposed dynamic multi-timestep adaptive embedding, global sub-interval deviation mapping, and multi-dimensional deviation compensation mechanisms jointly mitigate distortions and error accumulation caused by repeated latent-pixel transformations, ensuring stable and reliable secret extraction even at high embedding rates.

\subsection{Security}
In image steganography, security is typically categorized
into imperceptibility and anti-steganalysis performance.

\noindent\textbf{Imperceptibility.} Image quality serves as a primary metric for evaluating the imperceptibility of steganography. Table~\ref{tab:embedding_comparison} presents a comparison of different methods in terms of embedding rate and reconstruction error (MAE). DTAMS achieves an MAE of 0.5763 at a low embedding rate of 4.8 bpp and 1.2774 at a high embedding rate of 12 bpp. As shown in Table~\ref{tab:embedding_comparison}, DTAMS achieves relatively low reconstruction error even at high embedding rates, indicating that recovered secret images maintain high perceptual fidelity and color consistency.

\noindent\textbf{Anti-steganalysis Performance.} The detectability of stego images was evaluated using both conventional and deep-learning-based steganalysis methods, including YeNet \cite{ye2017deep}, SRNet \cite{8470101}, and SiaStegNet \cite{you2020siamese}. Table~\ref{tab:steganalysis} summarizes the detection accuracy of DTAMS in comparison with IDEAS, S2IRT, StegaDDPM, and LDStega across the FFHQ, Bedroom, and Cat datasets. DTAMS consistently achieves detection rates around 0.50 across all datasets and detectors, outperforming current SOTA methods.

\section{Conclusion}

This paper proposes a diffusion-based generative image steganography method, termed DTAMS, which enhances robustness and security under high embedding rates through dynamic multi-timestep adaptive embedding and multi-dimensional deviation compensation. By integrating a preprocessing-driven timestep selection strategy with sub-interval deviation mapping, DTAMS effectively suppresses distortions. Experimental results demonstrate that, at a high capacity of 12 bpp, DTAMS achieves over 99.2\% message extraction accuracy and maintains strong robustness and imperceptibility against various noise and compression attacks.

\bibliographystyle{named}
\bibliography{ijcai26}

@INPROCEEDINGS{Lugmayr2022repaint,
  author    = {Lugmayr, A. and Danelljan, M. and Romero, A. and Yu, F. and Timofte, R. and Van Gool, L.},
  title     = {RePaint: Inpainting Using Denoising Diffusion Probabilistic Models},
  booktitle = {Proceedings of the IEEE/CVF Conference on Computer Vision and Pattern Recognition (CVPR)},
  year      = {2022},
  pages     = {11461--11471}
}

@article{ye2017deep,
  title     = {Deep learning hierarchical representations for image steganalysis},
  author    = {Ye, J. and Ni, J. and Yi, Y.},
  journal   = {IEEE Transactions on Information Forensics and Security},
  volume    = {12},
  number    = {11},
  pages     = {2545--2557},
  year      = {2017}
}

@article{you2020siamese,
  title     = {A Siamese CNN for image steganalysis},
  author    = {You, W. and Zhang, H. and Zhao, X.},
  journal   = {IEEE Transactions on Information Forensics and Security},
  volume    = {16},
  pages     = {291--306},
  year      = {2020}
}

@inproceedings{Karras2019StyleGAN,
  author    = {T. Karras and S. Laine and T. Aila},
  title     = {A Style-Based Generator Architecture for Generative Adversarial Networks},
  booktitle = {Proceedings of the IEEE/CVF Conference on Computer Vision and Pattern Recognition (CVPR)},
  year      = {2019},
  pages     = {4401--4410}
}

@article{Yu2015LSUN,
  author  = {F. Yu and A. Seff and Y. Zhang and S. Song and T. Funkhouser and J. Xiao},
  title   = {LSUN: Construction of a Large-Scale Image Dataset Using Deep Learning with Humans in the Loop},
  journal = {ArXiv},
  year    = {2015},
  volume  = {1506.03365}
}

@article{Meng2025Video,
  author  = {Meng, L. and Jiang, X. and Xu, Q. and Sun, T.},
  title   = {A Robust Coverless Video Steganography Based on Two-Level DCT Features Against Video Attacks},
  journal = {IEEE Transactions on Multimedia},
  year    = {2025},
  volume  = {27},
  pages   = {6434--6448},
  doi     = {10.1109/TMM.2025.3586104}
}

@ARTICLE{Yan2025FGSAudioFF,
  author  = {Yan, J. and Cheng, Y. and Yin, Z. and Zhang, X. and Wang, S. and Sun, T. and Jiang, X.},
  title   = {FGS-Audio: Fixed-Decoder Framework for Audio Steganography with Adversarial Perturbation Generation},
  journal = {ArXiv},
  year    = {2025},
  volume  = {2505.22266}
}

@ARTICLE{cheng2025robust,
  author  = {Cheng, Y. and Luo, Z. and Yin, Z.},
  title   = {Robust steganography with boundary-preserving overflow alleviation and adaptive error correction},
  journal = {Expert Systems with Applications},
  year    = {2025},
  volume  = {281},
  pages   = {127598},
  doi     = {10.1016/j.eswa.2025.127598}
}

@ARTICLE{Cheng2025RFNNs,
  author  = {Cheng, Y. and Zhou, J. and Chen, J. and Yin, Z. and Zhang, X.},
  title   = {RFNNS: Robust Fixed Neural Network Steganography with Popular Deep Generative Models},
  journal = {Proceedings of the AAAI Conference on Artificial Intelligence}, 
  year    = {2026}
}

@ARTICLE{Hu2018DCGANStego,
  author  = {Hu, D. and Wang, L. and Jiang, W. and Zheng, S. and Li, B.},
  title   = {A Novel Image Steganography Method via Deep Convolutional Generative Adversarial Networks},
  journal = {IEEE Access},
  volume  = {6},
  pages   = {38303--38314},
  year    = {2018}
}

@INPROCEEDINGS{ho2020denoising,
  author    = {Ho, J. and Jain, A. and Abbeel, P.},
  title     = {Denoising Diffusion Probabilistic Models},
  booktitle = {Advances in Neural Information Processing Systems (NeurIPS)},
  year      = {2020},
  pages     = {6840--6851},
  url       = {https://proceedings.neurips.cc/paper/2020/file/4c5bcfec8584af0d967f1ab10179ca4b-Paper.pdf}
}

@INPROCEEDINGS{rombach2022high,
  author    = {Rombach, R. and Blattmann, A. and Lorenz, D. and Esser, P. and Ommer, B.},
  title     = {High-Resolution Image Synthesis with Latent Diffusion Models},
  booktitle = {Proceedings of the IEEE/CVF Conference on Computer Vision and Pattern Recognition (CVPR)},
  year      = {2022},
  pages     = {10684--10695},
  url       = {https://openaccess.thecvf.com/content/CVPR2022/html/Rombach_High-Resolution_Image_Synthesis_With_Latent_Diffusion_Models_CVPR_2022_paper.html}
}

@INPROCEEDINGS{Peng2023Stegaddpm,
  author    = {Peng, Y. and Hu, D. and Wang, Y. and Chen, K. and Pei, G. and Zhang, W.},
  title     = {Stegaddpm: Generative Image Steganography Based on Denoising Diffusion Probabilistic Model},
  booktitle = {Proceedings of the 31st ACM International Conference on Multimedia},
  pages     = {7143--7151},
  year      = {2023}
}

@ARTICLE{Yuan2025DGADM,
  author  = {Yuan, C. and Ji, Z. and Li, X. and Zhou, Z. and Xia, Z. and Wu, Q. J.},
  title   = {DGADM-GIS: Deterministic Guided Additive Diffusion Model for Generative Image Steganography},
  journal = {IEEE Transactions on Dependable and Secure Computing},
  year    = {2025}
}

@ARTICLE{Yu2023CROSS,
  author  = {Yu, J. and Zhang, X. and Xu, Y. and Zhang, J.},
  title   = {CROSS: Diffusion Model Makes Controllable, Robust and Secure Image Steganography},
  journal = {Advances in Neural Information Processing Systems (NeurIPS)},
  volume  = {36},
  pages   = {80730--80743},
  year    = {2023}
}

@INPROCEEDINGS{Peng2024LDStega,
  author    = {Peng, Y. and Wang, Y. and Hu, D. and Chen, K. and Rong, X. and Zhang, W.},
  title     = {LDStega: Practical and Robust Generative Image Steganography Based on Latent Diffusion Models},
  booktitle = {Proceedings of the 32nd ACM International Conference on Multimedia},
  pages     = {3001--3009},
  year      = {2024}
}

@INPROCEEDINGS{xu2025mddm,
  author    = {Xu, Z. and Xu, D. and Li, Z. and Zhang, C.},
  title     = {MDDM: Practical Message-Driven Generative Image Steganography Based on Diffusion Models},
  booktitle = {Forty-second International Conference on Machine Learning},
  year      = {2025}
}

@ARTICLE{Zhou2022SecretToImage,
  author  = {Zhou, Z. and Su, Y. and Li, J. and Yu, K. and Wu, Q. J. and Fu, Z. and Shi, Y.},
  title   = {Secret-to-Image Reversible Transformation for Generative Steganography},
  journal = {IEEE Transactions on Dependable and Secure Computing},
  volume  = {20},
  number  = {5},
  pages   = {4118--4134},
  year    = {2022}
}

@INPROCEEDINGS{Liu2022DisentanglementAE,
  author    = {Liu, X. and Ma, Z. and Ma, J. and Zhang, J. and Schaefer, G. and Fang, H.},
  title     = {Image Disentanglement Autoencoder for Steganography Without Embedding},
  booktitle = {Proceedings of the IEEE/CVF Conference on Computer Vision and Pattern Recognition (CVPR)},
  pages     = {2303--2312},
  year      = {2022}
}

@article{gao2026survey,
  title={A survey of fragile model watermarking},
  author={Gao, Z. and Cheng, Y. and Yin, Z.},
  journal={Signal Processing},
  volume={238},
  pages={110088},
  year={2026},
  publisher={Elsevier}
}

@article{zhou2025auto,
  title={Auto-Stega: An Agent-Driven System for Lifelong Strategy Evolution in LLM-Based Text Steganography},
  author={Zhou, J. and Cheng, Y. and Xie, Y. and Yin, Z.},
  journal = {ArXiv},
  year    = {2025},
  volume  = {2510.06565}
}

@inproceedings{long2025scf,
  title={SCF-Stega: Controllable Linguistic Steganography Based on Semantic Communications Framework},
  author={Long, Y. and Yang, Z. and Wang, Z. and Zhou, Z. and Huang, Y. and Zhou, L.},
  booktitle={ICASSP 2025-2025 IEEE International Conference on Acoustics, Speech and Signal Processing (ICASSP)},
  pages={1--5},
  year={2025}
}

@inproceedings{wu2024generative,
  title={Generative text steganography with large language model},
  author={Wu, J. and Wu, Z. and Xue, Y. and Wen, J. and Peng, W.},
  booktitle={Proceedings of the 32nd ACM International Conference on Multimedia},
  pages={10345--10353},
  year={2024}
}

@inproceedings{mao2024covert,
  title={From covert hiding to visual editing: robust generative video steganography},
  author={Mao, X. and Hu, X. and Peng, W. and Gan, Z. and Qian, Z. and Zhang, X. and Li, S.},
  booktitle={Proceedings of the 32nd ACM International Conference on Multimedia},
  pages={2757--2765},
  year={2024}
}

@article{luo2024comprehensive,
  title={A Comprehensive Survey of Digital Image Steganography and Steganalysis},
  author={Luo, W. and Wei, K. and Li, Q. and Ye, M. and Tan, S. and Tang, W. and Huang, J. and others},
  journal={APSIPA Transactions on Signal and Information Processing},
  volume={13},
  number={1},
  year={2024},
  publisher={Now Publishers, Inc.}
}

@article{guan2025non,
  title={Non-Binary Polar Codes for Steganography},
  author={Guan, Q. and Chen, K. and Lu, W. and Zhang, W. and Yu, N.},
  journal={IEEE Transactions on Dependable and Secure Computing},
  year={2025},
  publisher={IEEE}
}

@article{li2025coas,
  title={CoAS: Composite Audio Steganography Based on Text and Speech Synthesis},
  author={Li, Y. and Chen, K. and Wang, Y. and Zhang, X. and Wang, G. and Zhang, W. and Yu, N.},
  journal={IEEE Transactions on Information Forensics and Security},
  year={2025},
  publisher={IEEE}
}

@inproceedings{
song2021denoising,
title={Denoising Diffusion Implicit Models},
author={Song, J. and Meng, C. and Ermon, S.},
booktitle={International Conference on Learning Representations},
year={2021}
}

@ARTICLE{8470101,
  author={Boroumand, M. and Chen, M. and Fridrich, J.},
  journal={IEEE Transactions on Information Forensics and Security}, 
  title={Deep Residual Network for Steganalysis of Digital Images}, 
  year={2019},
  volume={14},
  number={5},
  pages={1181-1193}
}

@inproceedings{qi2025provably,
  title={Provably Secure Image Robust Steganography via Cross-modal Error Correction},
  author={Qi, Y. and Chen, K. and Zhao, N. and Yang, Z. and Zhang, W},
  booktitle={Proceedings of the AAAI Conference on Artificial Intelligence},
  volume={39},
  pages={26354--26362},
  year={2025}
}

@inproceedings{ma2023robust,
  title={Robust steganography without embedding based on secure container synthesis and iterative message recovery},
  author={Ma, Z. and Zhu, Y. and Luo, G. and Liu, X. and Schaefer, G. and Fang, H.},
  booktitle={IJCAI},
  pages={4838--4846},
  year={2023}
}

@inproceedings{yang2024diffstega,
  title={DiffStega: Towards Universal Training-Free Coverless Image Steganography with Diffusion Models},
  author={Yang, Y. and Liu, Z. and Jia, J. and Gao, Z. and Li, Y. and Sun, W. and Liu, X. and Zhai, G.},
  booktitle={IJCAI},
  year={2024}
}

@inproceedings{bao2024pluggable,
  title={Pluggable Watermarking of Deepfake Models for Deepfake Detection},
  author={Bao, H. and Zhang, X. and Wang, Q. and Liang, K. and Wang, Z. and Ji, S. and Chen, W.},
  booktitle={IJCAI},
  year={2024}
}

\end{document}